# Zero-day attack and ransomware detection

Steven Jabulani Nhlapo and Mike Nkongolo Wa Nkongolo

Department of Informatics, University of Pretoria, South Africa

u23963353@tuks.co.za; mike.wankongolo@up.ac.za

## Abstract

Zero-day and ransomware attacks continue to exploit computing infrastructure, revealing the limitations of traditional Network Intrusion Detection Systems (NIDS) in promptly classifying these threats. Despite efforts by cybersecurity specialists to implement preventive measures and reduce false positives and negatives, significant zero-day and ransomware attacks still occur annually. A modern solution is essential to address this issue effectively. Machine Learning (ML) models offer a promising approach to enhancing NIDS capabilities. This study utilises the *UGRansome* dataset to detect zero-day and ransomware attacks by training various ML models on its subsets. Results show that the Random Forest Classifier (RFC), XGBoost, and Ensemble Method achieved perfect scores across all performance metrics, including accuracy, precision, recall, and F1-score. In contrast, the Support Vector Machine (SVM) and Naïve Bayes (NB) models performed poorly, not reaching 100% in any metrics. When comparing these results with ransomware detection studies, a similar trend is observed. For instance, a study using the UGRansome dataset for ransomware detection highlighted that Decision Trees outperformed SVM and Multilayer Perceptron (MLP), achieving 98.83% accuracy, 99.41% precision, recall, and F-measure. Another study comparing six algorithms, including Decision Trees, RFC, and XGBoost, found these models highly effective, with accuracy scores around 99.4%, proving their suitability for real-time detection scenarios. Additionally, research on zero-day attacks using the Deep Forest approach with the UGRansome dataset achieved a 97.7% accuracy, outperforming SVM, Adaboost, and NB models. This aligns with the current study's findings, suggesting that Ensemble Methods and Decision Trees excel in capturing complex data relationships essential for cybersecurity applications. Future research should explore additional approaches, such as Synthetic Minority Over-sampling Technique (SMOTE), to improve model performance and consider diverse datasets and real-time detection methods to further bolster network defences against these evolving threats.

## 1. Introduction

Businesses have heavily invested in traditional Network Intrusion Detection Systems (NIDS) and complementary non-technical measures, including policy implementation, routine audits, and up-to-date vulnerability assessments [1]. While traditional NIDS play a crucial role in safeguarding organisational computing infrastructure, these systems often fall short against zero-day and ransomware attacks [1], [2]. Their limitations stem from an inability to promptly identify novel attack signatures or patterns, which can lead to prolonged undetected breaches impacting business operations, systems, and data. Machine Learning (ML)-based NIDS offer a promising alternative for zero-day and ransomware attack detection [3]. Unlike traditional systems, ML models are adept at identifying patterns and anomalies in network traffic that may signal zero-day vulnerabilities, even without prior knowledge of the attack's specifics [3], [4]. This capability allows ML-based NIDS to potentially detect and mitigate zero-day and ransomware threats before they are recognized and addressed by vendors. Given the potential damage these attacks can inflict on businesses, timely detection and prevention are critical [5]. Despite advancements, existing research on ML-enhanced NIDS often relies on outdated datasets such as KDD-CUP99, NSL-KDD, UNSW-NB15, and CICIDS-2017 [6]. These legacy datasets do not reflect the evolving nature of threats, which can undermine the effectiveness of ML models and lead to high false-positive rates. To overcome these limitations, this research uses the *UGRansome* dataset [9], which is employed to train and evaluate ML models for zero-day and ransomware attack recognition [10]. The study aims to evaluate the effectiveness of the UGRansome dataset in enhancing zero-day and ransomware attack detection when used to train ML models for NIDS by addressing the following objectives:

1. **Detection challenges:** Traditional NIDS often struggle to detect zero-day and ransomware vulnerabilities due to their novel signatures and patterns. This research explores how ML models can effectively classify various threat classes within the UGRansome dataset, which includes both normal and anomalous network activities. The goal is to determine whether this dataset can improve detection capabilities, particularly for previously unknown threats.
2. **Model performance:** The study assesses the performance of ML models when exposed to unknown data from the UGRansome dataset. This evaluation provides insights into the dataset's efficacy in enhancing the detection of zero-day and ransomware attacks.

The findings are based on a processed version of the UGRansome dataset, optimised for improved performance in zero-day attack recognition.

NIDS have relied on signature-based methods that use pre-defined rules to detect potential attacks [6], [7]. These systems, however, often fall short in identifying novel or unknown attacks. In contrast, ML-based NIDS can learn from historical data to detect both known and previously unseen threats without needing explicit rules [8]. This adaptive learning capability enhances their effectiveness in network security by automatically identifying and mitigating cyber-attacks. With this, the study aims to validate the UGRansome dataset as a valuable resource for training ML-based NIDS, specifically for improving the accuracy of zero-day and ransomware attack detection. By using this dataset, the study seeks to advance the development of more robust and efficient NIDS. The findings contribute to a more comprehensive defence against zero-day exploits, which can have severe financial and reputational repercussions for organisations. The use of the UGRansome dataset offers a novel perspective on the capabilities of ML-based NIDS, potentially leading to more sophisticated and precise detection systems. This research not only enhances the field of network security but also provides organisations with improved tools to safeguard against emerging cyber threats.

## 2. Background

The UGRansome dataset was created in 2021 by Nkongolo et al. [11]. The dataset is designed for detecting zero-day and ransomware attacks and anomalies that are not identifiable by existing threat signatures. This dataset captures modern network traffic and includes cyclostationary patterns for both normal and abnormal behaviours, with a focus on attacks that infiltrate network nodes. By incorporating features from real-life datasets such as UGR '16 and ransomware, UGRansome enhances the detection of Advanced Persistent Threats (APT), including zero-day threats like *Razy, Globe, EDA2, and TowerWeb*. Tested with cross-validation and compared to KDD99 and NSL-KDD datasets, it shows improved performance with minimal false alarms when using the Random Forest Classifier (RFC). Additionally, the challenge of balancing the dataset due to the non-uniform distribution of real-life threat classes was addressed in [11]. One year after its creation, Zahra and Sea [12] used a novel methodology for detecting intrusions in cloud computing environments by leveraging the UGRansome dataset. The proposed methodology incorporates multiple models—including RFC, Naive Baye (NB), Support Vector Machine (SVM), Ensemble Model Technique, and Genetic Algorithm Optimizer (GAO)—to enhance attack detection. Utilising cloud services such as *SageMaker* for execution and *S3 buckets* for dataset storage, the study evaluates model performance through metrics like confusion matrix, accuracy, F1 score, precision, and recall.

The results show that RFC achieved a 99% accuracy rate, while the Ensemble Model obtained 98%, outperforming other models in detecting the most recent attacks. In the same year, Tokmak [13] explored the application of a Deep Forest approach with a layered architecture for detecting zero-day attacks. To evaluate the effectiveness of this method, the research utilised both the UGRansome, and the widely used NSL-KDD datasets. Initial experiments with the UGRansome dataset achieved an accuracy of 97.7% using the Deep Forest algorithm. When compared to other ML algorithms such as SVM, Adaboost, NB, and Deep Neural Networks (DNN), the Deep Forest approach demonstrated superior performance. Subsequently, experiments with the NSL-KDD dataset yielded an impressive accuracy of 99.9% with the Deep Forest algorithm. Again, it outperformed the SVM, Adaboost, NB, and DNN methods, showcasing its effectiveness in both datasets. In 2024, Alhashmi et al. [14] used the UGRansome dataset for early detection of ransomware. The study assesses the performance of six ML algorithms for ransomware detection, including Logistic Regression, Decision Tree, NB, RFC, AdaBoost, and XGBoost. It evaluates these algorithms based on accuracy, precision, recall, and F1-score, as well as computational performance factors such as build time, training time, classification speed, and Kappa statistics [14]. The results indicate that RFC, Decision Tree, and XGBoost are particularly effective, with accuracy rates of 99.37%, 99.42%, and 99.48%, respectively. These algorithms also demonstrate high classification speed, making them suitable for real-time detection and capable of handling ransomware samples amid noise and data variations. Chaudhary and Adhikari [15] address the need for a comparative analysis of ML algorithms for ransomware detection using the UGRansome dataset. Key performance metrics including accuracy, precision, recall, and F-measure were evaluated through experiments. The results reveal that the Decision Tree algorithm outperforms SVM and Multi-Layer Perceptron (MLP) in all metrics, achieving an accuracy of 98.83%, precision of 99.41%, recall of 99.41%, and F-measure of 99.41%. In contrast, SVM and MLP demonstrated lower performance scores. The study highlights the Decision Tree's superior ability to capture non-linear data relationships, making it an effective model for ransomware detection. The primary contribution of [15] is in identifying the Decision Tree as a highly effective tool for detecting ransomware which significantly outperforms other ML models.

**Limitations**: Each study discussed in this section highlights strengths in specific areas but also has limitations [12], [13], [14], [15]. For instance, while accuracy rates are high, they may not fully account for the practical implementation challenges such as real-time detection and adaptability to new attack patterns (see Table 1).

Data cleaning and feature selection are crucial but may introduce biases or overlook subtle patterns in the data (Table 1). Additionally, the studies' methodologies may not be uniformly applicable across different datasets or evolving ransomware threats. Further research is needed to address these limitations and enhance the generalizability and robustness of the ML models (Table 1).

| Ref | ML model | Feature selection | ML algorithm | Accuracy | Limitation |
|---|---|---|---|---|---|
| [12] | Ensemble | GAO | RFC | 99% | Data cleaning |
| [13] | Layered architecture | - | Deep Forest | 99.9% | Feature selection |
| [14] | Ensemble | - | XGBoost | 99.48% | Feature selection |
| [15] | Comparative | - | Decision Tree | 98.83% | Data cleaning & feature selection |
| This study | Ensemble | - | RFC | 100% | Feature selection |

**Table 1.** Evaluation of studies using the UGRansome dataset

### 3. Methodology

An updated version of the UGRansome dataset was released on Kaggle in December 2023 [16], which is utilised in this research. The dataset includes various attack types such as DoS, Botnet, Neris Botnet, Spam, Blacklist, UDP Scan, and Port Scanning, along with 14 attributes across 149,043 data points. Figure 1 presents a detailed view of the features included in the UGRansome dataset [16] while Figure 2 illustrates the ML architecture.

**Data cleaning**

The dataset was reduced by eliminating seventy-three (73) entries associated with the attribute '1' in the 'ExpAddress' feature [16]. This enhancement improves the accuracy of threat identification compared to previous versions of the dataset. Additionally, the 'Time' feature has been updated to include positive 'Timestamp' attributes. Normalisation of the dataset is particularly important when managing large datasets, as it ensures that features with significant values, such as 'NetFlow_Bytes,' do not disproportionately affect other features like 'Time' [17] . The dataset reveals that the 'NetFlow_Bytes' attribute exhibits large values at the initial data points, which could potentially skew the 'Time' attribute when applied to ML models [18].

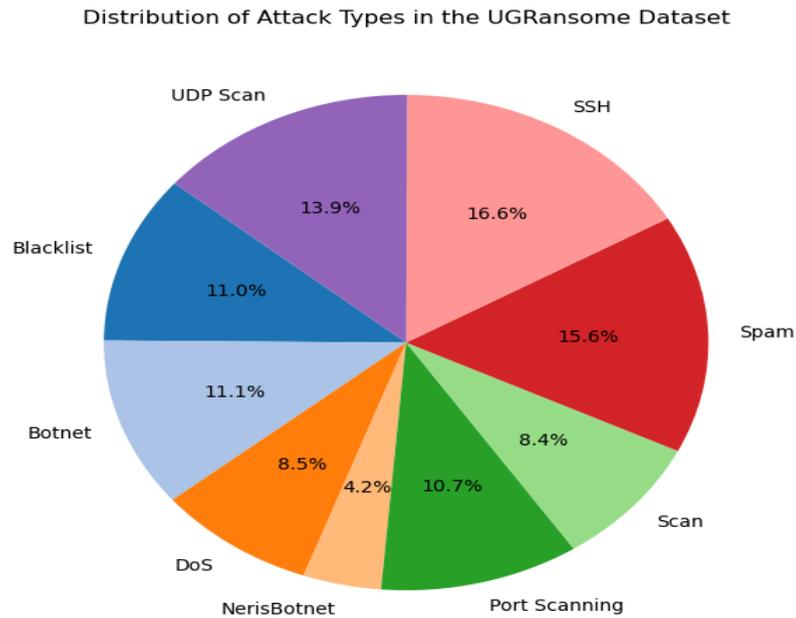

**Figure 1.** Attacks distribution in the UGRansome dataset

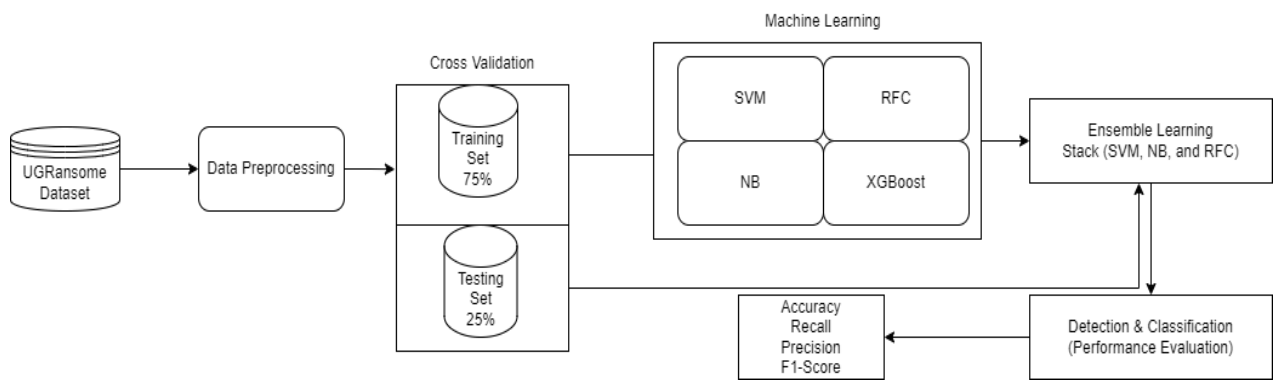

**Figure 2.** The proposed ML architecture

Normalisation mitigates the bias introduced by these large values and reduces the computational burden necessary for developing an effective ML model. Figure 3 illustrates the data distribution for the 'NetFlow_Bytes' feature before and after normalisation.

4. **Results**

Table 2 evaluates the SVM model's performance in classifying three classes: Anomaly (A), Signature (S), and Synthetic Signature (SS) from a test dataset of 29,794 samples. The overall accuracy of the SVM model is 69%, indicating that it correctly predicts 69% of the instances across all classes. For Anomaly (A), the model has a precision of 78% and a recall of 61%, resulting in an F1-score of 51%, highlighting a notable imbalance between precision and recall (Table 2).

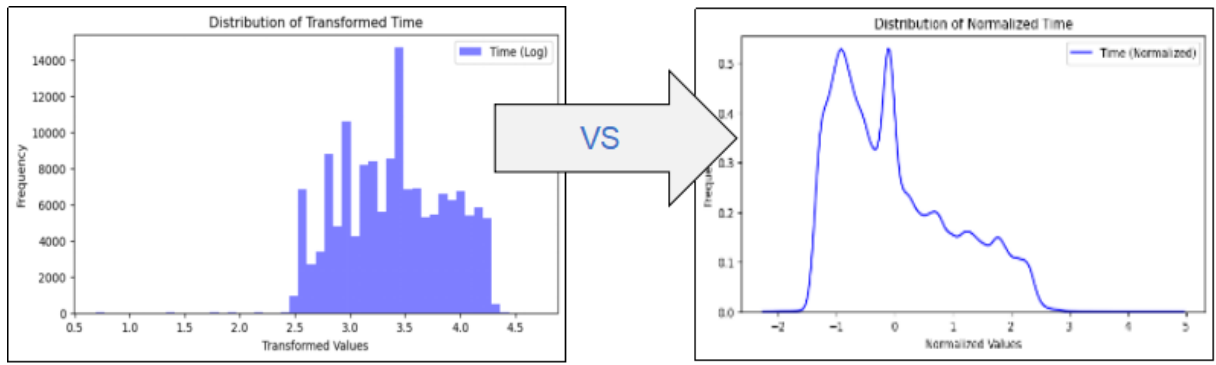

**Figure 3.** Data normalisation

| Class | Precision | Recall | F1-score | Test data |
|---|---|---|---|---|
| **0: A** | 0.78 | 0.61 | 0.59 | 7,375 |
| **1: S** | 0.63 | 0.79 | 0.70 | 13,608 |
| **2: SS** | 0.76 | 0.59 | 0.67 | 8,811 |
| **Accuracy** | | | *0.69* | *29,794* |

Table 2. The SVM results using the UGRansome dataset

For Signature (S), the precision is 63% and recall is 79%, with an improved F1-score of 70%, showing better balance and performance compared to Anomaly (A). For Synthetic Signature (SS), the model achieves a precision of 76% and a recall of 59%, leading to an F1-score of 67%, reflecting a moderate performance with some imbalance between true and false positives (Figure 4). The NB model was evaluated on three classes—Anomaly (A), Signature (S), and Synthetic Signature (SS)—using a test dataset of 29,794 entries. The model achieved an overall accuracy of 78% for Anomaly (A), with a precision of 92%, recall of 69%, and an F1-score of 79% (see Table 3). For Signature (S), with 12,077 occurrences, the precision was 71%, recall was 80%, and the F1-score was 75%. For Synthetic Signature (SS), which had 7,826 occurrences, the precision was 79%, recall was 89%, and the F1 score was 83% (Table 2).

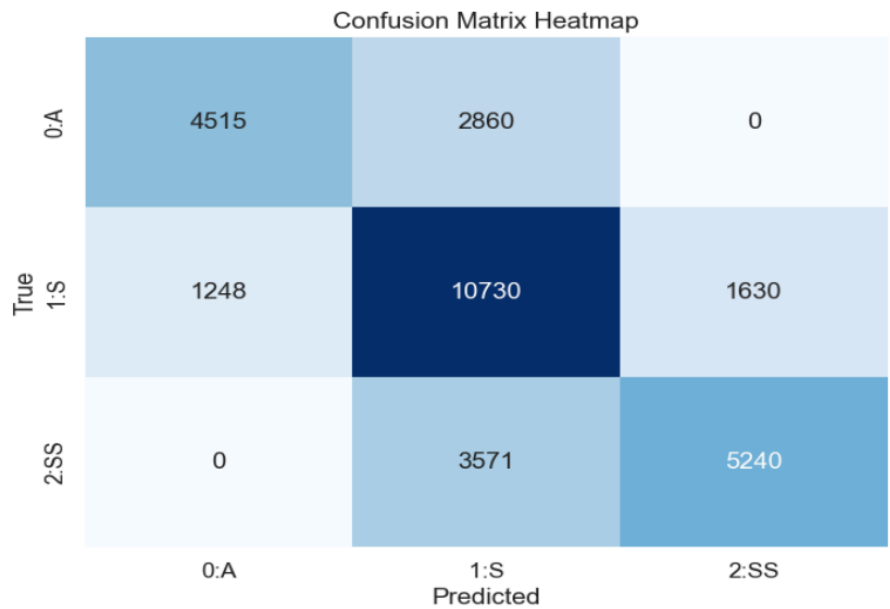

**Figure 4.** The SVM confusion matrix

| Class | Precision | Recall | F1-score | Test data |
|---|---|---|---|---|
| 0: A | 0.92 | 0.69 | 0.79 | 9,891 |
| 1: S | 0.71 | 0.80 | 0.75 | 12,077 |
| 2: SS | 0.79 | 0.89 | 0.83 | 7,826 |
| **Accuracy** | | | *0.78* | *29,794* |

**Table 3.** The NB results using the UGRansome dataset

The confusion matrix provides a detailed breakdown of the model's performance across these classes (Figure 5).

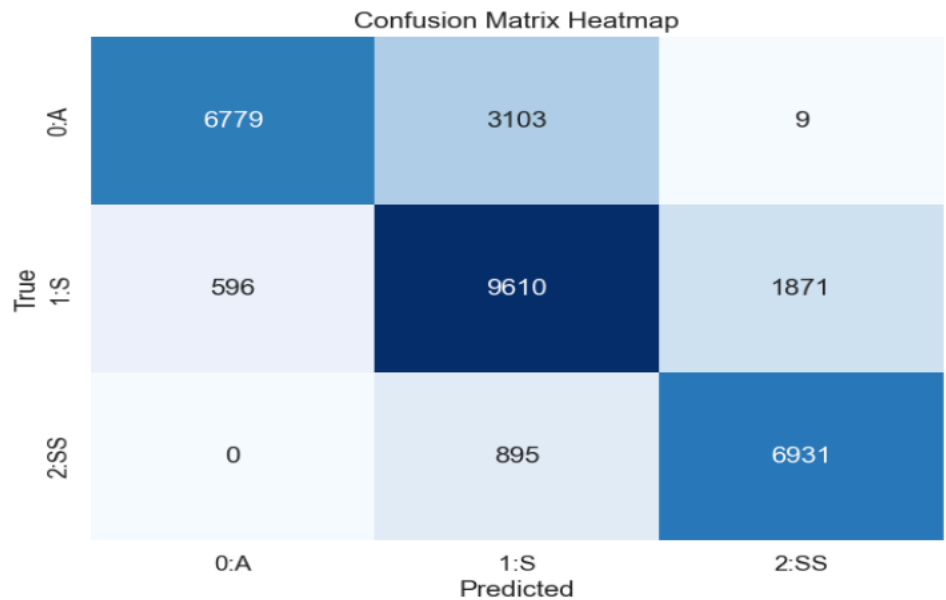

**Figure 5.** The NB confusion matrix

The RFC was evaluated on three classes, the model achieved a perfect accuracy of 100% for all classes (Table 4).

| Class | Precision | Recall | F1-score | Test data |
| --- | --- | --- | --- | --- |
| 0: A | 1.0 | 1.0 | 1.0 | 7,175 |
| 1: S | 1.0 | 1.0 | 1.0 | 13,608 |
| 2: SS | 1.0 | 1.0 | 1.0 | 8,811 |
| **Accuracy** | | | ***1.0*** | ***29,794*** |

**Table 4.** The RFC results using the UGRansome dataset

The confusion matrix provides a detailed numerical summary of these perfect predictions (Figure 6).

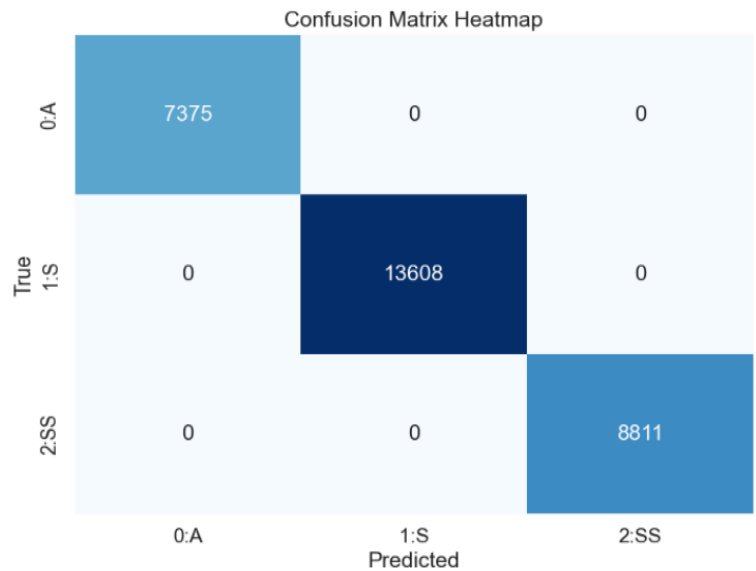

**Figure 6.** The RFC confusion matrix

| Class | Precision | Recall | F1-score | Test data |
|---|---|---|---|---|
| 0: A | 1.0 | 1.0 | 1.0 | 7,375 |
| 1: S | 1.0 | 1.0 | 1.0 | 13,608 |
| 2: SS | 1.0 | 1.0 | 1.0 | 8,811 |
| **Accuracy** | | | ***1.0*** | ***29,794*** |

**Table 5.** The XGBoost results using the UGRansome dataset

The XGBoost model achieved a perfect accuracy of 100% across all classes (Table 5). For Anomaly (A), the precision, recall, and F1-score were all 100%, indicating flawless detection.

Similarly, for Signature (S), with 13,608 occurrences, the model recorded 100% for all metrics (Table 5). For Synthetic Signature (SS), with 8,811 occurrences, the model also achieved 100% precision, recall, and F1-score. The confusion matrix illustrates these perfect predictions for each class (Figure 7). Figure 8 summarises the performance metrics of each model based on their accuracy scores.

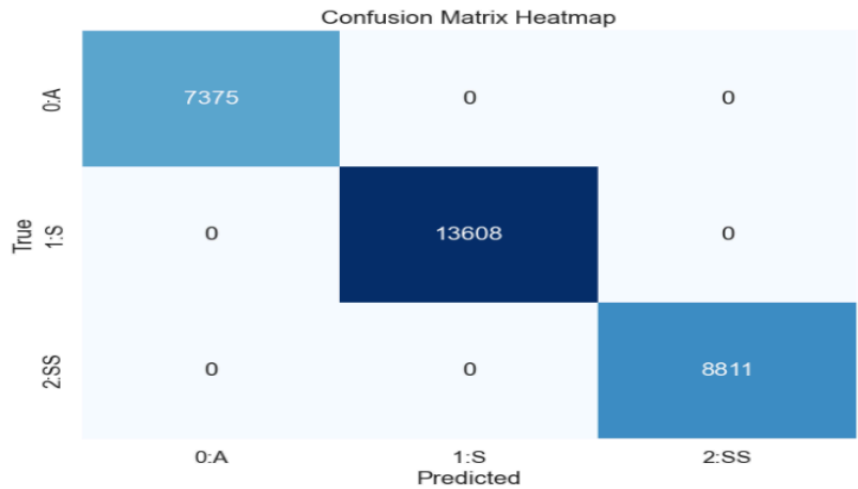

**Figure 7.** The XGBoost confusion matrix

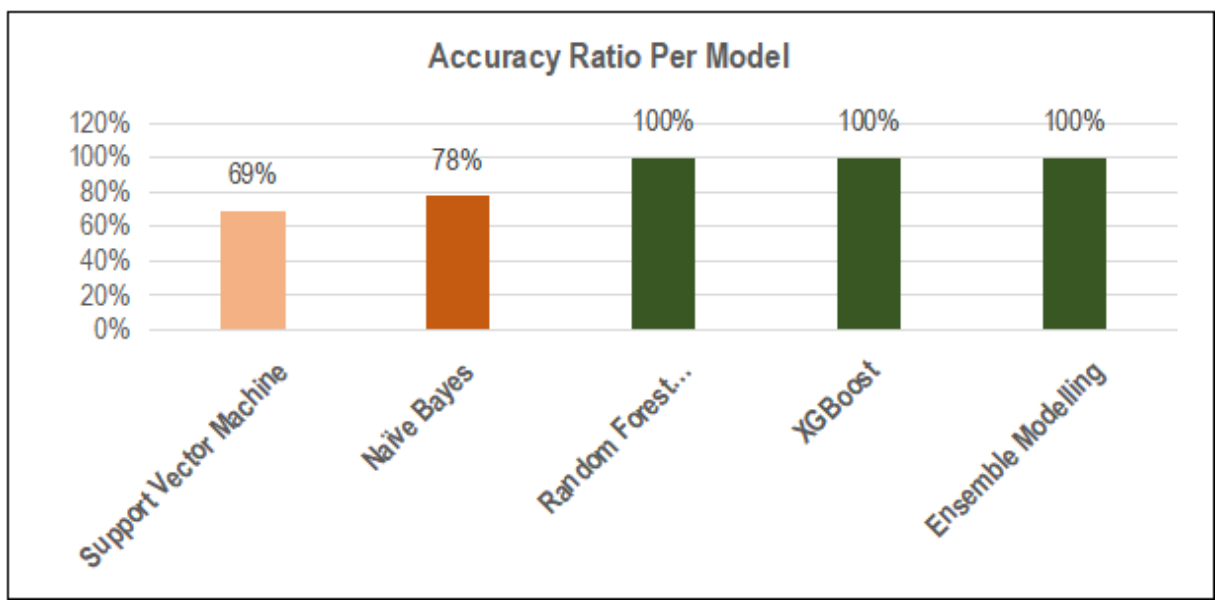

**Figure 8.** Experimental comparison of ML models

The Ensemble Method achieved a perfect accuracy of 100%, demonstrating the benefits of combining multiple models, specifically RFC and XGBoost, to enhance prediction robustness. In contrast, the SVM model and NB showed lower accuracies of 69% and 78%, respectively, indicating their reduced effectiveness (Figure 8).

The SVM model's performance was notably affected by a recall ratio of 0.59 for the Synthetic Signature (A) class, revealing challenges in identifying all relevant instances and a low F1-score of 0.59 for Anomaly (A). NBs had a 78% accuracy, with a precision ratio of 0.92 for Anomaly (A) but a lower recall ratio of 0.69, impacting its F1-score and overall performance.. The study's findings illustrate the substantial advantage of using Ensemble Methods for detecting zero-day and ransomware attacks (Figure 8). The Ensemble Method's perfect accuracy demonstrates that combining multiple ML models, such as RFC and XGBoost, significantly enhances detection capabilities by leveraging their individual strengths [19]. In contrast, the lower accuracy of models like SVM and NB indicates their limitations in recognizing novel threats [20], with SVM struggling with recall and NB showing high precision but moderate overall accuracy [21]. This highlights the necessity of integrating advanced and adaptive detection systems to address the complexities of new and evolving threats effectively. Future research should focus on refining these methods, exploring additional Ensemble Models, and utilising diverse datasets to improve detection rates. For practical applications, organisations should consider adopting ensemble-based systems to bolster their defences against sophisticated attacks [22], as relying solely on traditional or single-model approaches might not provide adequate protection.

5. **Conclusion**

Zero-day and ransomware attacks continue to exploit computing infrastructure, highlighting the limitations of traditional NIDS in effectively classifying these threats. Despite efforts to reduce false positives and negatives, significant attacks persist annually, necessitating modern solutions. ML models offer a promising approach to improving NIDS. This study, using the UGRansome dataset, finds that the RFC, XGBoost, and Ensemble Methods achieve perfect scores in accuracy, precision, recall, and F1-score. In contrast, SVM and NB models performed poorly. Similar trends are seen in other studies; for instance, Decision Trees significantly outperformed SVM and MLP in ransomware detection. Additionally, Deep Forest approaches achieved 97.7% accuracy in detecting zero-day attacks, surpassing other models.

**Dataset and code availability.** For access to the source code, please contact the author at the following email addresses: academia email (u23963353@tuks.co.za) or personal email (steven.jabulani@gmail.com). The revised dataset and source code can be found on GitHub and Kaggle.